\documentstyle[a4,11pt]{article} 

\input amssym.def       
\input amssym.tex

\def\goth{\frak}          
\def\double{\Bbb}

\def\cc{{\double C}}     
       
\def\zz{{\double Z}}

\def\rr{{\double R}}

\newtheorem{theorem}{Theorem}

\def\cp{\rtimes}

\def\Si{\Sigma}

\def\cinfc{C^{\infty}_c}
\newcommand{\be}{\begin{equation}}
\newcommand{\ee}{\end{equation}}
\newcommand{\beq}{\begin{eqnarray}}
\newcommand{\eeq}{\end{eqnarray}}
\newcommand{\om}{\omega}
\newcommand{\Om}{\Omega}
\newcommand{\al}{\alpha}

\newcommand{\non}{\nonumber}

\newcommand{\ch}{\mbox{ch}}

\newcommand{\Ac}{{\cal A}}

\newcommand{\te}{\theta}
\newcommand{\tb}{\overline{\theta}}

\def\zb{\overline{z}}
\def\Zb{\overline{Z}}

\def\d{\partial}
\def\db{\overline{\partial}}

\begin{document}

\begin{center}

{\large ON THE TOPOLOGICAL INTERPRETATION OF GRAVITATIONAL ANOMALIES}
\vskip 1cm
{\bf Denis PERROT\footnotemark[1]}
\vskip 0.5cm
Centre de Physique Th\'eorique, CNRS-Luminy,\\ Case 907, 
F-13288 Marseille cedex 9, France \\[2mm]
{\tt perrot@cpt.univ-mrs.fr}
\end{center}
\vskip 1cm
\begin{abstract} 
We consider the mixed gravitational-Yang-Mills anomaly as the coupling between 
the $K$-theory and $K$-homology of a $C^*$-algebra crossed product. The index 
theorem of Connes-Moscovici allows to compute the Chern character of the 
$K$-cycle by local formulae involving connections and curvatures. It gives a 
topological interpretation to the anomaly, in the sense of noncommutative 
algebras.
\end{abstract}

\vskip 1cm

\noindent {\bf MSC91:} 19D55, 81T13, 81T50\\

\noindent {\bf Keywords:} $K$-theory, cyclic cohomology, gauge theories.\\

\footnotetext[1]{Allocataire de recherche MENRT.}

\section{Introduction}

In a previous paper \cite{P} we proved a formula computing the topological 
anomaly of gauge theories, in the very general framework on noncommutative 
geometry \cite{C2}. This formula reduces just to the pairing between the 
$K$-theory classes of loops in the gauge group, and some $K$-homology classes 
arising from abstract Dirac-type operators. This simple remark allows one to 
compare the usual BRS machinery with cyclic cohomology \cite{C2}. Both are 
nontrivial as {\it local} cohomologies, but we feel that cyclic cohomology is 
more suitable since it can be directly related to the {\it analytic} content of 
the Dirac-type operator via the Chern character, whereas BRS cohomology has no 
obvious link with index theory in general.\\

In this paper we want to apply the topological anomaly formula in the mixed 
gravitational-Yang-Mills case, i.e. when the gauge group is the crossed product 
of Yang-Mills transformations on a manifold $X$ with a group of diffeomorphisms 
acting on $X$. Here the Chern character of the $K$-cycle involved takes its 
values in the cyclic cohomology of an algebra crossed product. The local index 
theorem of Connes-Moscovici \cite{CM98} then expresses this Chern character in 
terms of Gelfand-Fuchs cohomology. We shall compute it by connections and 
curvatures, and see that it gives expressions very similar to the usual ones 
encountered in the (BRS) study of gravitational anomalies. However there is an 
essential difference here: ordinary BRS methods deal with {\it Lie algebra} 
cohomology, whereas the characteristic classes for crossed products involve {\it 
group} cohomology. Of course both are related by van Est type theorems, but we 
insist on the fact that group considerations can describe gravitational 
anomalies topologically, as the pairing of nontrivial cyclic cocycles with the 
$K$-theory of a (noncommutative) algebra crossed product.\\

The paper is organized as follows. In section 2 we recall the anomaly formula 
in the case of Yang-Mills theories, with particular emphasis on its link with 
Bott periodicity, and we improve it by taking the diffeomorphisms into 
account.\\
In section 3 we present a relatively self-contained collection of some 
classical results of Bott and Haefliger \cite{Bo,H} concerning equivariant 
cohomology and Gelfand-Fuchs cohomology, and apply it to the Connes-Moscovici 
index theorem for crossed products.\\
In the last section we illustrate these tools by the study of conformal 
transformations on a Riemann surface. This gives rise to a nontrivial cyclic 
cocycle, corresponding to a conformal anomaly.


\section{The anomaly formula}

\subsection{Yang-Mills anomalies}

Let $X$ be an even-dimensional oriented smooth manifold, $C_0(X)$ the 
$C^*$-algebra of continuous complex-valued functions vanishing at infinity. We 
consider a $K$-cycle over $X$. For concreteness, let us take the signature 
complex: endow $X$ with a Riemannian metric and let $H=H_+\oplus H_-$ be the 
Hilbert space of $L^2$ differential forms on $X$, with $\zz_2$-graduation given 
by self- and anti-selfduality. The elliptic signature operator ${D}=d+d^*$ acts 
on a dense domain of $H$ as an odd unbounded selfadjoint operator. The pair 
$(H,{D})$ defines in this way a $K$-homology class $[{D}]\in K^*(C_0(X))$.\\
A typical situation in Qantum Field Theory is the following. Let $N$ be a 
positive integer, and consider the group $G=U_N(\cinfc(X))$ of $N\times N$ 
unitary matrices with entries in the algebra of smooth compactly supported 
functions $\cinfc(X)$. It is the group of Yang-Mills transformations, with 
structure group $U_N$ (if $X$ is not compact, one should add a unit). $G$ acts 
on the tensor product $H\otimes \cc^N$ by even bounded endomorphisms. In general 
the elliptic operator comes equipped with a Yang-Mills connection $A$,
\be
{D}_A={D}+A
\ee
which transforms under the gauge group according to the adjoint representation:
\beq
{D}_A&\rightarrow& u^{-1}{D}_Au\qquad\qquad u\in G\non\\
A&\rightarrow& u^{-1}Au+u^{-1}[{D},u]\ .
\eeq
As we work with $\zz_2$-graded spaces we adopt the usual matricial notation
\be
{D}_A=\left(\begin{array}{cc}
  0 & {D}_A^- \\
 {D}_A^+ & 0  \end{array}\right)\qquad
u=\left(\begin{array}{cc}
  u_+ & 0 \\
  0 & u_-  \end{array}\right)\qquad
H=\left(\begin{array}{c}
  H_+  \\
  H_-  \end{array}\right)\ .
\ee
Consider now the chiral action
\be
S(\psi_+,\psi_-,A)=\langle \psi_-,{D}_A^+\psi_+\rangle\qquad \psi_{\pm}\in 
H_{\pm}\ .
\ee
If we quantize the fields $\psi_{\pm}$ according to the Fermi statistics, then 
the vacuum functional
\be
Z(A)=\int [d\psi]\ e^{-S(\psi_+,\psi_-,A)}
\ee
is simply given by a regularized determinant $\det {D}_A^+$, see \cite{Si} (for 
the Bose statistics one takes the inverse determinant). In general it is not 
invariant under the gauge group, i.e.
\be
\det {D}_A^+ \neq \det (u^{-1}{D}_Au)^+\ .
\ee
Define the loop group $G^{S^1}$ as the set of smooth maps
\be
g: S^1\rightarrow U_N(\cinfc(X)) 
\ee
with base-point the identity. The product is pointwise. Let $t\in [0,2\pi)$ be 
the coordinate on $S^1$. Given a loop $g\in G^{S^1}$ the determinant $Z(t)=\det 
(g^{-1}(t){D}_Ag(t))^+$ is an invertible $\cc$-valued function on $S^1$ and the 
topological anomaly is just the winding number
\be
w={1\over{2\pi i}}\int_{S^1} {{d Z(t)}\over{Z(t)}}\quad\in\zz\ .
\ee
The anomaly formula of \cite{P} relates it to the $K$-cycle $[{D}]$ as follows. 
First the loop group $G^{S^1}$ may be identified with $U_N(\cinfc(S^1\times 
X))$. Then any element $g$ in the loop group determines a class $[g]$ of the 
$K$-theory group $K_1(C_0(S^1\times X))$. The operator
\be
Q=\left(\begin{array}{cc}
  i\d_t & {D}_A^- \\
 {D}_A^+ & -i\d_t  \end{array}\right) \label{k}
\ee
acting on sections of the Hilbert bundle $H\times S^1$, represents an odd 
$K$-cycle for the $C^*$-algebra $C_0(S^1\times X)$. Let $\ch_*(Q)$ be its Chern 
character \cite{C2} in the odd cyclic cohomology of $\cinfc(S^1\times X)$. One 
has a well-defined, homotopy-invariant, integral pairing
\be
w=\langle [g], \ch_*(Q)\rangle\ \in\zz\label{aa}
\ee
computing the value of the topological (Yang-Mills) anomaly on the loop $g$ 
\cite{P}.\\

Observe that in (\ref{aa}) any reference to the connection $A$ disappears. It is 
a purely topological formula involving the $K$-homology class of $Q$ and the 
$K$-theory element $[g]$. In particular if the first homotopy group of 
$U_N(C_0(X))$ is zero, then any loop $g$ is contractible and its image $[g]$ 
vanishes in $K_1(C_0(X\times S^1))$. By increasing $N$ we may eventually choose 
nontrivial loops detected by $Q$, which is really the essence of $K$-theory.\\
In \cite{P} we established the anomaly formula in a more general setting, 
allowing for $X$ to be a noncommutative ``space'' described by an associative 
algebra $\Ac$ together with an even Fredholm module $(H,D)$ \cite{C2} playing 
the role of the previous elliptic operator. Our goal in the following is to 
apply these ideas in the case of gravitational theories, where the gauge group 
contains the diffeomorphisms $\mbox{Diff}(X)$ of the manifold $X$. The relevant 
space is the groupoid $X\cp\mbox{Diff}(X)$, which is highly noncommutative in 
nature.\\

\subsection{The gravitational case}

We thus implement the above construction by taking the diffeomorphisms of $X$ 
into account. The group of mixed Yang-Mills $\cp$ gravitational transformations 
is the crossed product $U_N(\cinfc(X))\cp\mbox{Diff}(X)$, which lies in the 
matrix algebra of $\Ac=\cinfc(X)\cp\mbox{Diff}(X)$. The associative algebra 
$\Ac$ is generated by the symbols
\be
a=f U^*_{\psi}\qquad f\in\cinfc(X)\ ,\quad \psi\in\mbox{Diff}(X)\ ,
\ee
with product rule
\be
(f_1U^*_{\psi_1})(f_2U^*_{\psi_2})=f_1(f_2\circ\psi_1)U^*_{\psi_2\psi_1}\ ,
\ee
where $f_2\circ \psi_1$ is the pullback of $f_2$ by $\psi_1$. Since we are 
mostly concerned with $K$-theory, we shall enlarge this group to all invertible 
elements $Gl_N(\Ac)$ as well. Thus we are dealing with transformations involving 
matrices of diffeomorphisms. The physical interpretation seems obscure at first 
sight, but our motivation comes from the fact that the group $\mbox{Diff}(X)$ 
does not generally contain enough nontrivial loops. We shall see in the 
following that provided we consider matrix algebras, in the same philosophy as 
in the Yang-Mills case, the anomaly formula detects nontrivial topological 
objects related to diffeomorphisms. Let us explain carefully the construction in 
this case.\\
 
Let $\mbox{Diff}(S^1,X)$ denote the subgroup of $\mbox{Diff}(S^1\times X)$ 
consisting in diffeomorphisms $\psi$ such that
\be
pr\circ \psi =pr\ ,
\ee
where $pr:S^1\times X\rightarrow S^1$ is the projection onto the first factor. 
Then $\mbox{Diff}(S^1,X)$ plays the role of the loop group of $\mbox{Diff}(X)$. 
Thus we identify the loop group of $Gl_N(\Ac)$ with $Gl_N(\cinfc(S^1\times 
X)\cp\mbox{Diff}(S^1,X))$.\\

From now on put $M=S^1\times X$. For the sake of definiteness, let $\Gamma$ be a 
discrete countable subgroup of $\mbox{Diff}(S^1,X)$. We choose the loops as 
elements of $Gl_N(\cinfc(M)\cp\Gamma)$ and consider their images in the group 
$K_1$ of the $C^*$-algebra $C_0(M)\cp\Gamma$. As before we would like to 
evaluate these $K$-theory classes on some $K$-cycle $Q$. The previous signature 
operator is not suitable in this case because it does not define a $K$-cycle for 
$C_0(M)\cp\Gamma$. This problem is solved as in \cite{CM95} by passing to the 
bundle $P$ over $M$, whose fiber at $x\in M$ is the set of all euclidian metrics 
on the tangent space $T_xM$. $\Gamma$ acts canonically on $P$ and the $K$-theory 
of $C_0(M)\cp\Gamma$ lifts through the Thom map of \cite{C1}:
\be
\beta: \ K_*(C_0(M)\cp\Gamma)\rightarrow K_*(C_0(P)\cp\Gamma)\ .
\ee
On this bundle $P$ of metrics one can construct a differential operator $Q$ 
representing a $K$-cycle for $C_0(P)\cp\Gamma$, playing the role of the 
signature class \cite{CM95}. If we let $\ch_*(Q)$ be its Chern character in the 
cyclic cohomology of $\cinfc(P)\cp\Gamma$, the anomaly formula amounts to the 
computation of
\be
\langle \beta([g]), \ch_*(Q)\rangle\ ,\qquad [g]\in K_1(C_0(M)\cp\Gamma) 
\label{ano}
\ee
for any loop $g$. In the following sections we shall use the index theorem of 
Connes-Moscovici \cite{CM98} to express $\ch_*(Q)$ as an equivariant cohomology 
class. The latter is constructed from connections and curvatures in great 
analogy with the usual expressions of gravitational anomalies found in the 
literature. This together with the nontriviality of $\ch_*(Q)$ gives an 
interesting $K$-theoretical interpretation of these anomalies.\\

\subsection{Remark on Bott periodicity}

Note that in the context of $C^*$-algebras, the pure Yang-Mills anomaly has a 
simple interpretation in terms of Bott periodicity (\cite{Bl} \S 9). Indeed the 
set of homotopy classes of loops in $U_{\infty}(C_0(X))$ with base-point 1 is 
isomorphic to the group $K_1$ of the suspension of $C_0(X)$. Moreover the 
product in the loop group of $U_{\infty}(C_0(X))$ can eventually be taken as the 
concatenation of loops, so that 
\be
\pi_1(U_{\infty}(C_0(X)))\simeq K_1(C_0(X\times \rr))\ ,
\ee
and Bott periodicity stands for the isomorphism
\be
\te : \ K_0(C_0(X))\rightarrow K_1(C_0(X\times\rr))\ .
\ee
Also the Chern character of the differential operator (\ref{k}) is just the cup 
product
\be
\ch_*(Q)=\ch_*(D) \# [S^1]
\ee
between $\ch_*({D})$ in the cyclic cohomology of $\cinfc(X)$ and the fundamental 
class of the circle. Hence one has an equality (see e.g. \cite{C2} p. 225 prop. 
3 c))
\be
\langle [g], \ch_*(Q)\rangle =\langle \te^{-1}([g]),\ch_*({D})\rangle
\ee
for any loop $g\in G^{S^1}$. It follows that the evaluation of the Yang-Mills 
anomaly on a loop in the gauge group $U_{\infty}(\cinfc(X))$ is {\it equivalent} 
to the coupling between the $K$-theory of $X$ and the $K$-homology class $[D]$. 
This interpretation does not hold true for the gravitational anomaly because 
$C_0(M)\cp\Gamma$ is not the suspension of a $C^*$-algebra in general.

\section{Characteristic classes for crossed products}

In this section we recall basic facts about equivariant cohomology and 
Gelfand-Fuchs cohomology. Most of this material can be found in Bott-Haefliger 
\cite{Bo,H}. It allows to compute the characteristic classes of the crossed 
product $M\cp\Gamma$ appearing in the Connes-Moscovici index theorem 
\cite{CM98}, in terms of connections and curvatures.\\

\subsection{Equivariant cohomology}

Let $M$ be an oriented manifold, and $\Gamma$ a discrete group acting on $M$ by 
orientation-preserving diffeomorphisms. In the following we will not distinguish 
an element $g$ of $\Gamma$ and the corresponding diffeomorphism.\\
The space of homogeneous cochains of bidegree $n,m$ is zero if $n<0$ or $m<0$, 
otherwise it is the space $C^{n,m}(M)$ of maps $u$  from $\Gamma^{n+1}$ to the 
differential forms $\Om^m(M)$ of degree $m$ on $M$, subject to the equivariance 
condition
\be
u(g_0g,...,g_ng)=u(g_0,...,g_n)\circ g\ ,\qquad g_i,g\in\Gamma\ ,
\ee
where $\circ g$ denotes the pullback by the diffeomorphism $g$. On the complex 
$C^{*,*}$ one defines two differentials. The first one $\delta : 
C^{n,m}\rightarrow C^{n+1,m}$ is the simplicial differential
\be
(\delta u)(g_o,...,g_{n+1})= (-)^m\sum_{i=0}^{n+1}(-)^i 
u(g_0,...,\stackrel{\vee}{g_i},...,g_{n+1})\ ,
\ee
where $^{\vee}$ denotes omission. The second one $d:C^{n,m}\rightarrow 
C^{n,m+1}$ is the de Rham coboundary
\be
(du)(g_0,...,g_n)= d(u(g_0,...,g_n))\ .
\ee
The signs are chosen so that $d^2=\delta^2=d\delta+\delta d=0$. Geometrically, 
the total complex $(C^{*,*},d+\delta)$ describes the complex of cochains on the 
homotopy quotient $M_{\Gamma}=M\times_{\Gamma} E\Gamma$. By definition its 
cohomology is the {\it equivariant cohomology} $H^*(M_{\Gamma})$ of $M$.\\

It will be convenient for us to consider the following ring structure on 
homogeneous cochains. For $u\in C^{n,m}$ and $v\in C^{p,q}$, the product $uv\in 
C^{n+p,m+q}$ is 
\be
(uv)(g_0,...,g_{n+p})= (-)^{nq}u(g_0,...,g_n)v(g_n,...,g_{n+p})\ .
\ee
This product is associative and compatible with equivariance. Moreover the 
Leibniz rule is satisfied:
\beq
d(uv)&=& du\, v +(-)^{n+m}u\, dv\non\\
\delta(uv)&=& \delta u\, v+(-)^{n+m}u\, \delta v\ ,
\eeq
with $n+m$ the total degree of $u$. Thus $(C^{n,m},d+\delta)$ is a graded 
differential algebra.\\

Recall also \cite{H} that the above complex of homogeneous cochains is 
isomorphic to the complex of group cochains $C^*(\Gamma, \Om^*(M))$ with 
coefficients in the differential forms of $M$. To $u\in C^{n,m}(M)$ corresponds 
the group cochain $f\in C^n(\Gamma,\Om^m(M))$:
\be
f(g_1,...,g_n):= u(g_1...g_n,g_2...g_n,...,g_n,1)\ ,\label{b}
\ee
and the associated coboundary operator $\delta: C^n(\Gamma,\Om^m)\rightarrow 
C^{n+1}(\Gamma,\Om^m)$ reads
\beq
\delta f (g_1,...,g_{n+1})&=& f(g_2,...,g_{n+1}) 
+\sum_{i=1}^{n}(-)^if(g_1,...,g_ig_{i+1},...,g_{n+1})\non\\
&& + (-)^{n+1}f(g_1,...,g_n)\circ g_{n+1}\ .
\eeq

\subsection{Jet bundles}

Let $n$ be the dimension of the manifold $M$. Let $J^+_k$ be the space of 
$k$-jets of orientation-preserving diffeomorphisms 
\be
j: \rr^n\rightarrow M
\ee
from a neighborhood of $0$ in $\rr^n$ to $M$. Given a local coordinate system 
$\{x^{\mu}\}_{\mu=1,...,n}$ on $M$, $J^+_k$ has coordinates 
$\{x^{\mu},y^{\mu}_i,y^{\mu}_{i_1i_2},...,y^{\mu}_{i_1...i_k}\}$ corresponding 
to the jet
\be
j^{\mu}(u)= x^{\mu}+ y^{\mu}_i u^i + {1\over 2}y^{\mu}_{ij}u^iu^j 
+\cdot\cdot\cdot + {1\over{k!}}y^{\mu}_{i_1...i_k}u^{i_1}...u^{i_k}\ ,\qquad 
u\in\rr^n\ .
\ee
In particular the matrix $(y^{\mu}_i)$ belongs to $Gl^+_n(\rr)$, and the real 
numbers $y^{\mu}_{i_1...i_l}$ are symmetric in low indices.\\

$J^+_k$ is a principal bundle over $M$ with structure group $G^k$ consisting in 
the set of $k$-jets $h$ fixing $0$: $h^{\mu}(0)=0$. The right action of $G^k$ on 
$J^+_k$ is simply the composition of jets:
\be
j\rightarrow j\circ h\ ,\qquad j\in J^+_k\ ,\ h\in G^k\ .
\ee
Since any $(k+1)$-jet yields a $k$-jet, $J^+_{k+1}$ is a principal bundle over 
$J^+_k$ with structure group the kernel of the projection $G^{k+1}\rightarrow 
G^k$. We get in this way a tower of bundles
\be
\cdot\cdot\cdot \rightarrow J^+_k\rightarrow \cdot\cdot\cdot\rightarrow J^+_1 
\rightarrow M\ .
\ee
We write the inverse limit $J^+_{\infty}$. Note that $J^+_1$ is the bundle of 
oriented frames on $M$.\\

The action of $\Gamma$ on $M$ lifts on $J^+_k$ by left composition of jets
\be
j\rightarrow g\circ j\ ,\qquad j\in J^+_k\ ,\ g\in\Gamma
\ee
and clearly commutes with $G^k$. In particular the group $SO_n\subset Gl^+_n$ 
sits in $G^k$ as a maximal compact subgroup and $\Gamma$ still acts on the 
quotient $P_k=J^+_k/SO_n$, which is a bundle with contractible fiber over $M$. 
The action of an element $a\in SO_n$ is given by the right composition by the 
jet
\be
h^i(u)=a^i_j u^j\ ,\qquad u\in \rr^n\ ,
\ee
where $a=(a^i_j)$ is a matrix in $SO_n$. Explicitly the vertical coordinates of 
a point in $J^+_k$ change according to the rule
\be
y^{\mu}_{i_1...i_l}\rightarrow y^{\mu}_{j_1...j_l}a^{j_1}_{i_1}...a^{j_l}_{i_l}\ 
.
\ee
Thus one has a tower of bundles with contractible fiber
\be
\cdot\cdot\cdot \rightarrow P_k\rightarrow \cdot\cdot\cdot\rightarrow P_1 
\rightarrow M\ ,
\ee
with inverse limit $P_{\infty}$. Remark that $P_1$ is the bundle of metrics over 
$M$. Since $\Gamma$ lifts on each $P_k$, the homotopy quotient 
$P_{k,\Gamma}=P_k\times_{\Gamma}E\Gamma$ is a bundle over $M_\Gamma$ with 
contractible fiber, which induces an isomorphism in equivariant cohomology,
\be
H^*(P_{k,\Gamma})\simeq H^*(M_{\Gamma})\ ,
\ee
and also for the limit $H^*(P_{\infty,\Gamma})$.\\

\subsection{Gelfand-Fuchs cohomology}

Given a coordinate chart $\{x^{\mu},y^{\mu}_i,...,y^{\mu}_{i_1...i_k},...\}$, we 
identify {\it locally} $J^+_{\infty}$ with the pseudogroup of all 
diffeomorphisms $\rr^n\rightarrow \rr^n$. Its Lie algebra $\goth{a}$ corresponds 
to the formal vector fields of $\rr^n$. Let $\Om_{inv}^*(J^+_{\infty})$ be the 
complex of invariant forms under the left action of Diff$(M)$ on jets by 
composition
\be
j\rightarrow \varphi\circ j\ ,\qquad \varphi\in \mbox{Diff}(M)\ .
\ee
It is naturally isomorphic to the complex of Lie algebra cochains 
$C^*(\goth{a},\rr)$. An algebraic basis of $\mbox{Diff}(M)$-invariant forms on 
$J^+_{\infty}$ is provided by expanding the Maurer-Cartan form ``$j^{-1}\circ 
dj$'' in powers of $u\in\rr^n$:
\be
(j^{-1}\circ dj)^i(u)= \te^i+\te^i_j u^i+ {1\over 2}\te^i_{jk}u^ju^k 
+...+{1\over{k!}}\te^i_{j_1...j_k}u^{j_1}...u^{j_k}+... \ .
\ee
Due to the $\mbox{Diff}(M)$-invariance, the one-forms $\te$ are globally defined 
on $J^+_{\infty}$. Actually $\te^i_{j_1...j_k}$ is already defined on 
$J^+_{k+1}$. For example
\be
\te^i= (y^{-1})^i_{\mu}dx^{\mu}
\ee
lies on $J^+_1$ (here $((y^{-1})^i_{\mu})$ is the inverse matrix of 
$(y_i^{\mu})$),
\be
\te^i_j= (y^{-1})^i_{\mu}dy^{\mu}_j- 
y^{\mu}_{jk}(y^{-1})^i_{\mu}(y^{-1})^k_{\nu}dx^{\nu}
\ee
lies on $J^+_2$, and so on. Thus the Gelfand-Fuchs cohomology 
$H^*(\goth{a},\rr)$ is naturally isomorphic to the cohomology of invariant forms 
$H^*(\Om_{inv}^*(J^+_{\infty}))$. It is computed as follows \cite{G}. The group 
$Gl_n(\rr)$ acts on $\rr^n$ be linear diffeomorphisms. Let 
$\goth{g}\subset\goth{a}$ be its the Lie algebra. The Weil algebra associated to 
$\goth{g}$ is the tensor product
\be
W=\wedge \goth{g}^*\otimes S(\goth{g}^*)
\ee
of the exterior algebra on the dual space $\goth{g}^*$ of $\goth{g}$, by the 
symmetric algebra $S(\goth{g}^*)$. $W$ is a graded differential algebra: it is 
generated by the elements of degree 1
\be
\om^i_j\in \wedge^1\goth{g}^*
\ee
of the canonical basis of $\goth{g}^*$ and
\be
\Om^i_j\in S^1(\goth{g}^*)
\ee
of degree 2. A differential $d_W$ is uniquely defined by
\beq
d_W\om^i_j&=& \Om^i_j-\om^i_k\om^k_j\ ,\non\\
d_W\Om^i_j&=& \Om^i_k\om^k_j-\om^i_k\Om^k_j\ .
\eeq
Next we consider $\te^i_j$ as the coefficients of a connection 1-form on 
$\goth{a}$ with values in $\goth{g}$, and its curvature
\be
R^i_j = d\te^i_j+\te^i_k\te^k_j\ .
\ee
Then from Chern-Weil theory one has a morphism
\be
\psi: W\rightarrow \Om^*_{inv}(J^+_{\infty})
\ee
which sends $\om^i_j$ onto $\te^i_j$ and $\Om^i_j$ onto $R^i_j$. Furthermore the 
2-form $R^i_j$ is proportional to $dx^{\mu}$ and hence any polynomial in $R$ of 
degree $>n$ vanishes. It follows that $\psi$ factorises through $W_n$, the 
quotient of $W$ by the differential ideal generated by the elements in 
$S(\goth{g}^*)$ of degree $>2n$. The first result of Gelfand-Fuchs is \cite{G}

\begin{theorem}
The map $\psi: W_n\rightarrow \Om^*_{inv}(J^+_{\infty})$ induces an isomorphism 
in cohomology.
\end{theorem}

This theorem also admits a version relative to the action of $SO_n$ on 
$\goth{a}$. The complex of $SO_n$-basic cochains $C^*(\goth{a},SO_n)$ is 
naturaly isomorphic to the invariant forms on $P_{\infty}=J^+_{\infty}/SO_n$. 
Since $SO_n\subset Gl_n$, $W_n$ is a $SO_n$-algebra. Let $WSO_n$ be its 
subalgebra of basic elements relative to the action of $SO_n$. Then $\psi$ maps 
$WSO_n$ to $\Om^*_{inv}(P_{\infty})$ and one has \cite{G}

\begin{theorem}
The map $\psi$ induces an isomorphism
\be
H^*(WSO_n)\simeq H^*(\Om^*_{inv}(P_{\infty}))\ .
\ee
\end{theorem}

Next we want to send these classes into the equivariant cohomology of $M$. 
Remark that there is an injection
\be
i: \Omega^m(P_{\infty})\rightarrow C^{0,m}(P_{\infty})\ ,
\ee
which to any (non necessarily invariant) differential form $\al$ on $P_{\infty}$ 
associates the homogeneous 0-cochain
\be
\al(g_0):= \al\circ g_0\ ,\qquad \forall g_0\in\Gamma\ .
\ee
It is clear that under this map the image of a closed form in 
$\Om^*_{inv}(P_{\infty})$ is both $d$- and $\delta$-closed, and hence defines an 
equivariant cohomology class on $P_{\infty,\Gamma}$. Thus one gets a canonical 
map
\be
H^*(WSO_n)\rightarrow H^*(P_{\infty,\Gamma})\simeq H^*(M_{\Gamma})\ .
\ee
Note finally that the image of $WSO_n$ by $\psi$ lives in $P_2=J^+_2/SO_n$ since 
the forms $\te^i_j$ and $R^i_j=d\te^i_j+\te^i_k\te^k_j$ are defined on $J^+_2$. 
It is then sufficient to work on $P_2$ instead of $P_{\infty}$.\\

\subsection{Computation of $H^*(WSO_n)$}

We restrict to the case of a manifold $M$ of {\it odd} dimension $n$. In the 
truncated Weil algebra $W_n$, the Chern classes $c_i$, $i=1...n$, correspond to 
the terms of degree $2i$ in the determinant of the $n\times n$ matrix $1+\Om$. 
In particular:
\be
c_1=\Om^i_i\ ,\qquad c_2={1\over 2}((\Om^i_i)^2-\Om^i_j\Om^j_i)\ ,\qquad 
c_n=\det\Om\ .
\ee
For $i$ odd one can choose an element $u_i$ of degree $2i-1$ in $WSO_n$ such 
that $d_Wu_i=c_i$. Let $E(u_1,u_3,...,u_n)$ be the exterior algebra in the 
$u_i$, $i$ odd $\le n$, and $\rr[c_1,c_2,...,c_n]$ the algebra of polynomials in 
all the $c_i$ quotiented by the ideal of elements of degree strictly higher than 
$2n$. The tensor product
\be
WU_n= \rr[c_1,...,c_n]\otimes E(u_1,...,u_n)
\ee
is endowed with the differential $d$ such that $du_i=c_i$. Then one has \cite{G}

\begin{theorem}
The inclusion $WU_n\rightarrow WSO_n$ induces an isomorphism in cohomology.
\end{theorem}

In particular if we define the Pontrjagin classes
\be
p_i=c_{2i}\qquad \forall i\le {{n-1}\over 2}\ , \ n\ \mbox{odd}
\ee
then $H^*(WSO_n)$ always contains the polynomial algebra 
$\rr[p_1,p_2,...]_{trunc}$ in the $p_i$'s truncated by the elements of degree 
$>2n$.\\

\subsection{The Connes-Moscovici index theorem}

Let $P=P_1$ be the bundle of metrics over the odd-dimensional manifold $M$. On 
$P$ the hypoelliptic signature operator $Q$ of \cite{CM95} defines a $K$-cycle 
for the algebra $C_0(P)\cp\Gamma$. By \cite{C2} $\S$ III.2.$\delta$ one has an 
injective map
\be
\Phi: H^*(P\times_{\Gamma}E\Gamma)\hookrightarrow HC^*_{per}(\cinfc(P)\cp\Gamma)
\ee
from equivariant cohomology to the periodic cyclic cohomology of the crossed 
product $\cinfc(P)\cp\Gamma$. The index theorem of \cite{CM98} states that the 
Chern character $\ch_*(Q)\in HC^*_{per}(\cinfc(P)\cp\Gamma)$ is in the range of 
Gelfand-Fuchs cohomology. Actually $\ch_*(Q)$ has a preimage in the Pontrjagin 
ring $\rr[p_1,p_2...]_{trunc}$. \\
If we apply this construction to the situation of section 2, where $M=S^1\times 
X$ and $\Gamma$ is a loop group of diffeomorphisms on $X$, the complete 
computation of the anomaly formula (\ref{ano})
\be
\langle \beta([g]), \ch_*(Q)\rangle\qquad [g]\in K_1(C_0(M)\cp\Gamma)
\ee
yields an expression containing the image of the Pontrjagin classes in 
$H^*(M_{\Gamma})$ and other characteristic classes accounting for the Thom 
isomorphism $\beta$. In the following section we compute the image of the 
Pontrjagin ring in the particular case of Riemann surfaces and conformal 
transformations, and see that the result looks like familiar gravitational 
anomalies. The same holds clearly true in the general case.

\section{Application to Riemann surfaces}

Let us have a look at the simplest example. We take $M$ as the product of $S^1$ 
by a Riemann surface $\Si$. We view it as a trivial fiber bundle over $S^1$ with 
fiber $\Si$. Let $\Gamma$ be a discrete (pseudo)group of orientation preserving 
diffeomorphisms on $M$ fulfiling the two conditions\\

\noindent i) Each fiber $\Si$ over $S^1$ is globally $\Gamma$-invariant,\\
\noindent ii) The restriction of $\Gamma$ to a fiber is a conformal 
transformation of $\Si$.\\

Thus according to section 2 an element of $\Gamma$ is a loop of conformal 
transformations of $\Si$. Choose a local coordinate system $(z,\zb)$ related to 
the complex structure of $\Si$, and let $t\in [0,2\pi)$ be the variable on 
$S^1$. For any $g\in\Gamma$ we write
\be
z\circ g= Z_g\ ,\qquad \zb\circ g=\Zb_g\ ,\qquad t\circ g=t\ .
\ee
The jet bundles $J^+_k$ have local coordinates 
$(x^{\mu},y^{\mu}_i,...,y^{\mu}_{i_1...i_k})$, where the indices $\mu,i_l...$ 
can assume any of the three values $(z,\zb,t)$. Of course $x^{\mu}$ are 
identified with the coordinates on $M$:
\be
x^z=z\ ,\qquad x^{\zb}=\zb\ ,\qquad x^t=t\ .
\ee
Since the (real) dimension of $M$ is $n=3$ the Pontrjagin ring of the 
Gelfand-Fuchs cohomology $H^*(WSO_3)$ only contains the unit 1 and the first 
Pontrjagin class $p_1=c_2$. From the last section we know that $p_1$ is 
represented by a closed $\Gamma$-invariant 4-form on the bundle 
$P_2=J^+_2/SO_3$, explicitly given in terms of the tautological curvature 
$R^i_j$, $i,j=(z,\zb,t)$:
\be
p_1={1\over 2}((R^i_i)^2-R^i_jR^j_i)\quad \in\Om^4_{inv}(P_2)\ .\label{a}
\ee
We denote $\hat{p}_1$ its image in $H^4(P_{2,\Gamma})\simeq H^4(M_{\Gamma})$.\\

\subsection{Restriction to a subbundle}

Since $\Gamma$ is a group of conformal transformations of the fibers $\Si$ 
leaving $t$ invariant, one can restrict the geometry to the subbundle 
$\tilde{J}^+_2$ of $J^+_2$ consisting in holomorphic 2-jets
\be
u\in \rr^3\rightarrow j(u)\in M\ ,
\ee
which read in coordinates
\beq
j^z(u)&=&z+y^z_zu^z+y^z_tu^t+ {1\over 2}y^z_{zz}u^zu^z+y^z_{zt}u^zu^t+ {1\over 
2}y^z_{tt}u^tu^t\ ,\non\\
j^{\zb}(u)&=&{\zb}+y^{\zb}_{\zb}u^{\zb}+y^{\zb}_tu^t+ {1\over 
2}y^{\zb}_{{\zb}{\zb}}u^{\zb}u^{\zb}+y^{\zb}_{{\zb}t}u^{\zb}u^t+ {1\over 
2}y^{\zb}_{tt}u^tu^t\ ,\non\\
j^t(u)&=& t+u^t\ .
\eeq
Then the 2-jets of the elements of $\Gamma$ are contained in $\tilde{J}^+_2$. 
$\tilde{J}^+_2$ is a principal bundle over $M$, whose structure group contains 
$SO_2$ as a maximal compact subgroup. The action of $SO_2$ is obtained by the 
right composition
\be
j\in \tilde{J}^+_2\rightarrow j\circ h\in\tilde{J}^+_2\ ,
\ee
where $h$ is the jet of the rotation by an angle $\al$:
\be
h^z(u)=e^{i\al}u^z\ ,\qquad h^{\zb}(u)=e^{-i\al}u^{\zb}\ ,\qquad h^t(u)=u^t\ .
\ee
Thus $\tilde{P}_2=\tilde{J}^+_2/SO_2$ is a $\Gamma$-bundle over $M$ with 
contractible fiber so that $H^*(\tilde{P}_{2,\Gamma})=H^*(M_{\Gamma})$. Moreover 
$\tilde{P}_2$ is a $\Gamma$-invariant subbundle of $P_2$ and the injection 
$\tilde{P}_2\rightarrow P_2$ is a homotopy equivalence. Now $\hat{p}_1\in 
H^*(M_{\Gamma})$ may equivalently be represented by a closed invariant form on 
$\tilde{P}_2$ corresponding to the pullback of (\ref{a}). One computes that the 
pullbacks of the curvature coefficients $R^i_j$ are nonzero only for $R^z_z, 
R^z_t, R^{\zb}_{\zb}, R^{\zb}_t$, hence
\be
\hat{p}_1= {1\over 2} ((R^z_z+R^{\zb}_{\zb})^2-(R^z_z)^2-(R^{\zb}_{\zb})^2) = 
R^z_z R^{\zb}_{\zb}
\ee
is the pullback of $\hat{p}_1$ on $\tilde{J}^+_2$, and is $SO_2$-basic. In terms 
of the tautological connection $\te^i_j$ (eq. (\ref{b})) on $\tilde{J}^+_2$ one 
has $R^z_z=d\te^z_z$, with
\be
\te^z_z= (y^{-1})^z_zdy^z_z- y^z_{zz}(y^{-1})^z_z((y^{-1})^z_zdz+(y^{-1})^z_tdt) 
- y^z_{zt}(y^{-1})^z_z dt\ ,
\ee
and similarly for $R^{\zb}_{\zb}$. In the following we shall write $R$ (resp. 
$\overline{R}$) instead of $R^z_z$ (resp. $R^{\zb}_{\zb}$) and $\te$ (resp. 
$\tb$) instead of $\te^z_z$ (resp. $\te^{\zb}_{\zb}$). Remark that the 1-form 
$\te+\tb$ is $SO_2$-basic, which implies that the cohomology class of 
$R+\overline{R}$ in the $\Gamma$-invariant forms on $\tilde{P}_2$ is zero. Thus 
$R\overline{R}$ is cohomologous to $-R^2$ and we shall keep the latter as a 
representative of $\hat{p}_1$.\\

It is possible now to express $\hat{p}_1$ as an equivariant cocycle on 
$M_{\Gamma}$. Choose a K\"ahler metric $\rho(z,\zb)dz\otimes d\zb$ on $\Si$. 
Then the associated connection on $\tilde{J}^+_2$ is the globally defined (not 
$\Gamma$-invariant) 1-form
\be
\om = (y^{-1})^z_zdy^z_z + dz\partial_z \ln\rho\ .
\ee
Of course it corresponds to the $^z_z$ component of the connection form 
associated with $\rho$ on the frame bundle. We shall regard it as an equivariant 
cochain on $\tilde{J}^+_2$ through the inclusion
$\Om^1(\tilde{J}^+_2)\rightarrow C^{0,1}(\tilde{J}^+_2)$. The equivariant 
curvature $\Om=(d+\delta)\om$ is an element of $C^{0,2}(\tilde{J}^+_2)\oplus 
C^{1,1}(\tilde{J}^+_2)$:
\beq
\Om(g_0,g_1)&=& \delta\om(g_0,g_1)= -\om\circ g_1 +\om\circ g_0\ ,\non\\
\Om(g_0)&=& d\om(g_0)=d\om\circ g_0\ ,\qquad g_i\in\Gamma\ .
\eeq
In fact $\Om$ lives in $M_{\Gamma}$. Indeed, for $g_i\in\Gamma$, let $Z'_i$ 
denote the function $\partial_z(z\circ g_i)$. One has (with $\d=dz\d_z$)
\be
\om\circ g_i = (y^{-1})^z_zdy^z_z +d\ln Z'_i + (\partial\ln\rho)\circ g_i\ ,
\ee
so that
\beq
\Om(g_0,g_1)&=& d\ln Z'_0 -d\ln Z'_1 + (\partial\ln\rho)\circ g_0 
-(\partial\ln\rho)\circ g_1\ ,\non\\
\Om(g_0)&=& -(\partial\db\ln\rho)\circ g_0\ .
\eeq
Here $\d\db\ln\rho$ is the curvature 2-form of the K\"ahler metric. Using the 
multiplicative structure on equivariant cohomology (section 3) we consider the 
cocycle $-\Om^2$. It is cohomologous to $-R^2$ in $H^4(\tilde{P}_{2,\Gamma})$, 
indeed:
\be
\Om^2-R^2 = {1\over 2}(d+\delta)((\om-\te)(\Om+R)+ (\Om+R)(\om-\te))\ ,
\ee
and $\om-\te$ is a $SO_2$-basic equivariant 1-form on $\tilde{J}^+_2$. Thus we 
have proved

\begin{theorem}
The equivariant 4-cocycle $-\Om^2$ represents the image of $p_1\in H^*(WSO_3)$ 
in $H^4(M_{\Gamma})$. 
\end{theorem}

\subsection{Link with conformal anomalies}

Using formula (\ref{b}) we can express $\hat{p}_1$ as a group cocycle 
$\tilde{p}_1$ in $C^1(\Gamma,\Om^3(M))\oplus C^2(\Gamma,\Om^2(M))$: 
\be
\tilde{p}_1(g)=\hat{p}_1(g,1)\ ,\qquad 
\tilde{p}_1(g_1,g_2)=\hat{p}_1(g_1g_2,g_2,1)\ .
\ee
The first component $\tilde{p}_1(g)$ is related to conformal anomalies as 
follows. Let $g:S^1\rightarrow \mbox{Diff}(\Si)$ be a loop of conformal 
transformations of $\Si$, that is, $g\in \mbox{Diff}(S^1,\Si)$ according to the 
notations of section 2. Then $\tilde{p}_1(g)$ is a 3-form on $M=S^1\times\Si$: 
\beq
\tilde{p}_1(g)&=& -\Om(g,1)\Om(1)-\Om(g)\Om(g,1)\non\\
&=& (d\ln Z'+(\d\ln\rho)\circ g)R_{\rho}+\non\\
&&+R_{\rho}\circ g((d\ln Z')\circ g^{-1}-(\d\ln\rho)\circ g^{-1})\circ g\ ,
\eeq
where $Z=z\circ g$ and $Z'=\d_zZ$. $R_{\rho}=\d\db\ln\rho$ is the curvature 
associated to $\rho$. Let us define the $z$-component of the ghost vector field
\be
\xi^z = dt\d_t Z\circ g^{-1}\ .
\ee
It is a one-form on $S^1$ with values in the (conformal) vector fields of $\Si$. 
Equivalently it is the pullback of the Maurer-Cartan form on $\mbox{Diff}(\Si)$ 
by the loop $g$. One computes that
\be
R_{\rho}(d\ln Z'+(\d\ln\rho)\circ g)= R_{\rho}\,(D_z\xi^z)\circ g\ ,
\ee
where $D_z\xi^z=\d_z\xi^z+\xi^z\d_z\ln\rho$ is the covariant derivative. In the 
same way define
\be
(\xi^{-1})^z=dt\d_t (z\circ g^{-1})\circ g\ ,
\ee
one has
\be
(R_{\rho}\circ g)((d\ln Z')\circ g^{-1}-(\d\ln\rho)\circ g^{-1})\circ g= 
-(R_{\rho}\circ g) D_z(\xi^{-1})^z\ .
\ee
If the loop $g$ is the identity of $\Si$ at $t=0$, then 
\be
\tilde{p}_1(g)_{t=0}=2R_{\rho}D_z\xi^z \label{var}
\ee
is the usual expression for the infinitesimal variation, under the ghost vector 
field $\xi$, of the vacuum functional of a field theory on $\Si$, i.e. a 
gravitational anomaly.\\ 
Now the Chern character of the signature operator $Q$ (section 2), contains the 
image of $\tilde{p}_1$ by the injection
\be
\Phi: H^*(M_{\Gamma})\simeq H^*(P\times_{\Gamma}E\Gamma)\hookrightarrow 
HC^*(\cinfc(P)\cp\Gamma)\ ,
\ee
where $P$ is the bundle of all metrics (not necessarily K\"ahler) on the 
3-dimensional {\it real} manifold $M$. The topological anomaly formula then 
gives an {\it integrated} version of the infinitesimal variation (\ref{var}), 
and is in general nonzero, provided we evaluate the anomaly on invertible {\it 
matrices} over the algebra $\cinfc(M)\cp\Gamma$.\\

\subsection{Non-triviality of $\tilde{p}_1$}

To show that $\tilde{p}_1$, and consequently its image in 
$HC^*(\cinfc(P)\cp\Gamma)$, is in general a non-trivial cohomology class, we 
shall construct a cycle $c$ in the equivariant homology with compact support 
$H_*(M_{\Gamma})$, whose evaluation on $\tilde{p}_1$ is nonzero. Since it is 
sufficient to do this in a particular case, let us take for $\Si$ the Riemann 
sphere $\cc\cup \{\infty\}$, and $\rho(z,\zb)=1$. Then the only nonzero 
component of $\tilde{p}_1$ lies in $C^2(\Gamma,\Om^2(M))$:
\beq
\tilde{p}_1(g_1,g_2)&=&-\Om^2(g_1g_2,g_2,1)\ =\ \Om(g_1g_2,g_2)\Om(g_2,1)\non\\
&=& (d\ln Z'_1)\circ g_2\, d\ln Z'_2\ .
\eeq
The equivariant homology is computed by the bicomplex $(C_{n,m})_{n,m\ge 0}$,
\be
C_{n,m}= \cc[\Gamma]^{\otimes n}\otimes\Om_m(M)\ ,
\ee
where $\cc[\Gamma]$ is the group ring of $\Gamma$ and $\Om_m(M)$ the space of 
$m$-dimensional de Rham currents with compact support on $M$. The first boundary 
map $\delta:C_{n,m}\rightarrow C_{n-1,m}$ is
\beq
\delta (g_1\otimes...\otimes g_n\otimes C)&=&g_2\otimes..\otimes g_n\otimes C 
+\non\\&&+\sum_{i=1}^n (-)^i g_1\otimes...\otimes g_ig_{i+1}\otimes...\otimes 
g_n\otimes C+\non\\
&& + (-)^{n+1}g_1\otimes...\otimes g_{n-1}\otimes g_nC\ ,
\eeq
where $g_nC$ is the left action of $g_n\in\Gamma$ on the current $C\in\Om_m(M)$ 
by pushforward. The second differential $\d:C_{n,m}\rightarrow C_{n,m-1}$ is the 
de Rham boundary (not to be confused with the previous $dz\d_z$ !)
\be
\d(g_1\otimes...\otimes g_n\otimes C)=(-)^n g_1\otimes...\otimes g_n\otimes \d 
C\ .
\ee
We shall construct the cycle $c$ as an element of $C_{1,3}\oplus C_{2,2}$. Let 
$\Gamma$ be such that $g_1,g_2\in\Gamma$ with
\be
z\circ g_j=Z_j={{e^{itn_j}}\over z}\ ,\qquad t\circ g_j=t\ ,\qquad j=1,2\ ,\quad 
 n_j\in\zz\ .
\ee
Choose an orientation on $M=S^1\times \Si$ and let $C\in\Om_3(M)$ be the current 
corresponding to the integration of 3-forms over the full cylinder
\be
C=\{(z,\zb,t)\in M | z\zb\le 1\}\ .
\ee
One checks that
\be
g_i\d C=-\d C\ ,\qquad g_1g_2 C=C\ ,
\ee
which implies that
\be
c:= g_1\otimes g_2\otimes \d C + (g_2-g_1-g_1g_2)\otimes C
\ee
represents an homology class in $H_4(M_{\Gamma};\zz)$:
\be
(\d+\delta)c=0\ .
\ee
Therefore the pairing between $\tilde{p}_1$ and $c$ is simply given by
\be
\langle \tilde{p}_1, c\rangle = \int_{\d C}\tilde{p}_1(g_1,g_2)\ ,
\ee
which gives, up to an irrelevant sign depending on the orientation, the 
difference $8\pi^2(n_1-n_2)$.

\vskip 1cm

\noindent{\bf Acknowledgments:} The author wishes to thank S. Lazzarini for his 
constant support, and S. Majid for comments.



\begin{thebibliography}{9}


\bibitem{Bl} Blackadar B.: {\it $K$-theory for operator algebras}, 
Springer-Verlag, New-York (1986).

\bibitem{Bo} Bott R.: On characteristic classes in the framework of Gelfand-Fuks 
cohomology, Soci\'et\'e Math\'ematique de France, {\it Ast\'erisque} {\bf 32-33} 
(1976).

\bibitem{C1} Connes A.: Cyclic cohomology and the transverse fundamental class 
of a foliation. In: Geometric methods in operator algebras, Kyoto (1983), pp. 
52-144, Pitman Res. Notes in Math. 123 Longman, Harlow (1986).

\bibitem{C2} Connes A.: {\it Non-commutative geometry}, Academic Press, New-York 
(1994).


\bibitem{CM95} Connes A., Moscovici H.: The local index formula in 
non-commutative geometry, {\bf GAFA 5} (1995) 174-243.

\bibitem{CM98} Connes A., Moscovici H.: Hopf algebras, cyclic cohomology and the 
transverse index theorem, {\it Comm. Math. Phys.} {\bf 198} (1998) 199-246.


\bibitem{G} Godbillon C.: Cohomologies d'alg\`ebres de Lie de champs de vecteurs 
formels, S\'eminaire Bourbaki, vol. 1972/73, n$^o$ 421.

\bibitem{H} Haefliger A.: Differentiable cohomology, C.I.M.E. (1976).

\bibitem{P} Perrot D.: BRS cohomology and the Chern character in noncommutative 
geometry, preprint math-ph/9910044, to appear in {\it Lett. Math. Phys.}

\bibitem{Si} Singer I. M.: Families of Dirac operators with application to 
physics, Soc. Math. de France, {\it Ast\'erisque}, hors s\'erie (1985) 323-340.


\end{thebibliography}
\end{document}